\documentstyle[12pt]{article}
\newcommand{\be}{\begin{eqnarray}}
\newcommand{\ee}{\end{eqnarray}}
\newcommand{\add}{\addtocounter{equation}{-1}}
\newcommand{\no}{\nonumber}
\begin{document}

\begin{titlepage}
\vspace*{3.0cm}
\begin{center}
{\large\bf The Super $W_3$ Conformal algebra\\
and the Boussinesq hierarchy}
\vskip 0.5 cm
by
\vskip 0.5 cm
Ziemowit Popowicz\renewcommand{\thefootnote}{\dagger}\footnote{
Permanenent address: Institute of Theoretical Physics, University of Wroclaw
pl. M. Borna 9, 50-205 Wroclaw, Poland. e-mail ZIEMEK@PLWRUW11}\\
Research Institute for Theoretical Physics, P.O.
Box 9 (Siltavuorenpenger 20 C), FIN-00014 University of Helsinki, Finland.
\end{center}
\vskip 1.5 cm
{\bf Abstract}: The bihamiltonian structure of the $N=2$ Supersymmetric
Boussinesq equation is found. It is not reduced to the corresponding classical
structure and hence it describes the pure supersymmetric effect. For the
supersymmetric
Boussinesq equation which contains the classical partner the Lax pair is given
explicitly. Thus we prove the integrability of this equation.
\end{titlepage}

\section{Introduction}
Th Boussinesq equation
$$u_{tt}=-\gamma^2(u_{xxx}+8uu_x)_x\eqno{(1.1)}$$
with $\gamma$ an arbitrary constant, was first derived to describe
shallow-water
waves propagating in both directions [1]. Its equivalent form is
$$\left(\begin{array}{c}u \\ v\end{array}\right)_t=\gamma\left(\begin{array}{l}
v_x\\-u_{xxx}-8uu_x\end{array}\right)\ .\eqno{(1.2)}$$
which appears to be more comfortable for the Lax formulation. Indeed, the Lax
operator
$$L=\partial_{xxx}+2u\partial_x+v\ ,\eqno{(1.3)}$$
characterizes the $Bsq$ hierarchy via the Lax equations
$$L_{tn}=\gamma\biggl[ (L^{n/3})_+,L\biggr]\ ;\hspace{1.5cm}n=1,2,...
\eqno{(1.4)}$$
where $\partial_x=\partial/\partial_x$ and + denotes the projection of the
fractional
power $L^{n/3}$ onto its purely differential part. The operators $L$ and
$L^{n/3}$ in the formula (4) we shall call the Lax pair.

The Boussinesq equation itself corresponds to the first non-trivial flow
given by $n=2$. The system described by the Boussinesq equation is the
bi-hamiltonian system. The first hamiltonian structure was found by Gelfand
and Dickey [2] while the second one was conjectured and constructed by Adler
[3]. The proof of the validity of this conjecture can be found in [4].

In the 1988 several scientists [5-7] recognized that the second hamiltonian
structure of the Boussineq hierarchy, the Gelfand-Dickey algebra [8]
associated to the Boussinesq Lax operator, is a classical representation of
the so called $W$-algebra.

The $W$-algebras were first introduced by Zamolodchikov [9] in order to
extend polynomially the Virasoro algebra by higher spin fields. Since the
introduction of $W$-algberas they have been the subject of intensive
investigation by both physicists and mathematicians. The unexpected
relationship
between the $W$-algebras and the noncritical string [10] and Toda lattice
models [11] has been found.

On the other hand, in the last three years there has been a considerable
interest in the supersymmetric extension of Zamolodchikov's $W_n$
algebras. Recently the super $W$-algebra has been constructed in the
extended and non-extended case [12-16]. The tempting problem with possibly
important implication in the super $W_3$ gravity and the related matrix models,
is to find the generalized super Boussinesq-type hierarchy via the Lax
pair approach. The similar program has been succesfully applied to the
supersymmetric Korteweg-de Vries equation which is connected with the
supersymmetric extension of the $W_2$ algfebra (e.g. with the super Virasoro
algbera) [17-19].

There has been some attempt to utilize the Lax operator for the construction
of the nonextend super $W_3$ algebra as well as for the construction of
the supersymmetric extension of the Boussinesq equation. For example Nam
[20] examined a generalization of the Miura transformation and accompanying
factorization of the even Lax operator. However, he concludes that the
nontrivial factorization of the Lax operator is impossible and thus a
nontrivial supersymmetric version of the extended conformal algebra cannot
be constructed. On the other hand Figueroa et al [13] constructed the
non-extended supersymmetric version of the $W_3$ algebra using the odd Lax
operator. However, his odd Lax operator does not generate the equation of
motion and hence there is no the supersymmetric version of the
Boussinesq equation in his approach. The negative results in this framework
have been
reported also in [21,22].

Recently Ivanov and Krinovos [16] have constructed the nontrivial hamiltonian
flow on the extended super $W_3$ algebra yielding $N=2$ superextension of
the Boussinesq equation. Their superextension turns out to involve a free
parameter and is reducible in the bosonic sector to the Boussinesq equation
only at a special value of this parameter. The super $W_3$ algebra has been
obtained by the supersymmetrization of the Miura transformation and the
existencce of such a transformation is the important property  of the given
system. Indeed, such a transformation relates the second Poisson bracket
with the generalized Gardner-Zakharov-Faddeev bracket [8]. Moreover such
a transformation allows us to construct the representation of the $N=2$ super
$W_3$ algebra out of the superfield of the lower  conformed dimension than
the one characterizing the original fields.

The appearence of the Miura transformation is a strong indication that the
given system
could possesses the bihamiltonian structure. In this paper we show that indeed
for the special values of a free parameter of the supersymmetric
Boussinesq equation this equation has bihamiltonian structure. Interestingly
for this value of a free parameter the supersymmetric Boussinesq equation
does not contain the classical Boussinesq equation.

For the equation which contains the classical partner we have found the Lax
formulation using the symbolic manipulation package REDUCE [23]. It appeared
that the roots of the Lax operator are not uniquely determined, hence we check
all possibilities allowing for the existence of the Lax pair and finally
we arrive to the three different Lax formulations.
\section{The Bi-hamiltonian structure of the Super symmetric Boussinesq
equation}
The classical Boussinesq equation (1.2) can be written down as a bi-Hamiltonian
system
$$\left(\begin{array}{c}u \\ v\end{array}\right)_t=P_1\bigtriangledown\cdot
\int \biggl(\frac{1}{2}u^2_x-\frac{4}{3}u^3+\frac{1}{2}v^2\biggr) dx=P_2
\bigtriangledown\cdot\int\frac{1}{2}vdx\eqno{(2.1)}$$
Here $\bigtriangledown=\left(\begin{array}{c}\delta/\delta u\\\delta/\delta u
\end{array}\right)$ and the two Hamiltonian operators $P_1$ and $P_2$ are given
by
$$P_1=\left(\begin{array}{cc}0 & \partial\\ \partial & 0\end{array}\right)
\eqno{(2.2)}$$
$$P_2=\left(\begin{array}{lll}\partial^3+u\partial+\partial u & , & 2\partial
v+v\partial\\ \partial v+2v\partial & , &
-(\partial^5+2(\partial^3u+u\partial^3
)+3(\partial^2u\partial+\partial u\partial^2)\\ & & +8(\partial u^2+u^2
\partial))
\end{array}\right)\ .\eqno{(2.3)}$$
The Poisson bracket defined by the Hamiltonian operator $P_2$ corresponds
to the classical version of the $W_3$ algebra. In order to obtain the $P_2$
one can use the Miura transformation in the form
$$u=P_{1x}-\frac{1}{2}(P^2_1+P^2_2)\eqno{(2.4)}$$
$$v=-P_{2xx}+3P_1P_{2x}+P_{1x}P_2+\frac{2}{3}P^3_2-2P^2_1P_2\eqno{(2.5)}$$
and assume the following Poisson bracket (free field represention)
on the fields $P_1$ and $P_2$
$$\{ P_i,P_j\}=-\delta_{ij}\delta(x-y)\ .\eqno{(2.6)}$$
The compatibility of the two Hamiltonian structures is granted by the
observation that the simply shift $v\rightarrow v+\lambda$ produces the
Hamiltonian operator $P_2+3\lambda P_1$.

Now we would like to check wheather the similar properties occur in the
supersymmetric level. The basic object in the supersymmetric analysis is the
superfield and the sypersymmetric derivative. The superfields are the
superfermions or the superbosons depending, in addition to $x$ and $t$, upon
two anticommuting variables, $\theta$, and $\theta_2(\theta_1\theta_2=-
\theta_2\theta_1\ ,\ \theta^2_1=0\ ,\ \theta^2_2=0)$. Its Taylor
expansion with respect to the $\theta$'s is
$$\phi(x,\theta_1\theta_2)=\omega(x)+\theta_1\xi_1(x)+\theta_2\xi_2(x)+
\theta_2\theta_1u(x)\eqno{(2.7)}$$
where in below the fields $\omega,v$ are to be interpreted as the boson
(fermion) fields while $\xi_1,\xi_2$ as the fermion (boson) fileds for the
superboson (superfermions) field $\phi$ respectivelly. We choose the following
representaion on the superderivatives
$$D_1=\partial_{\theta_1}-\frac{1}{2}\theta_2\partial_x\ \ ,\ \ \
D_2=\partial_{\theta_2}-\frac{1}{2}\theta_1\partial_x\eqno{(2.8)}$$
and as the consequence we obtain
$$D_1D_1=D_2D_2=0\eqno{(2.9)}$$
$$\{ D_1D_2\}=-\partial_x\ .\eqno{(2.10)}$$
In the paper [16] author construct the supersymmetric operator product
expansion
(SOPE) of the $N=2$ super $W_3$ algebra in terms of the spin 1 and spin 2
supercurrent. This representaion has been obtained by the free superfield
realization or in other terms via the super Miura transformation. In order
to transform SOPE to the Poisson bracket we use the supersymmetric version
of the Cauchy theorem
$$\frac{1}{2\pi i}\int dz_1(z_{12})^{-n-1}(\theta_{12}\bar{\theta}_{12},
\theta_{12},\bar{\theta}_{12},1)=$$
$$=\frac{1}{n!}(1,D_1,-D_2,\frac{1}{2}[D_1D_2])(\partial^n\Phi)
\eqno{(2.11)}$$
where
$$\theta_{12}=\theta_1-\theta_2\ \ ,\ \ \ \bar{\theta}_{12}=\theta_1'-\theta_2'
\ \ ,\ \ \ z_{12}=z_1-z_2+\frac{1}{2}(\theta_1\theta_2'-\theta_2\theta_1')\ .$$
As the result we obtain the super analog of the $P_2$ operator with the
following entries
\addtocounter{equation}{+1}
\renewcommand{\theequation}{\arabic{equation}.12}
\be
P_{2_{11}}& =& \frac{c}{4}[{\cal D}_1,{\cal D}_2]\partial_x+{\cal J}_x+{\cal
J}\partial_x+\no\\
& & +({\cal D}_1{\cal J}){\cal D}_2+({\cal D}_2{\cal J}){\cal D}_1\no\\
P_{2_{12}}&=&2\cdot\partial_x\cdot T+(D_2T)D_1+(D_1T)D_2\no\\
P_{2_{21}}&=&\partial\cdot T+T\partial_x+(D_2T)D_1+(D_1T)D_2\no\\
P_{2_{22}}&=&-\frac{c}{8}[D_1,D_2]\partial_{xxx}-2{\cal J}\partial_{xxx}-6
({\cal D}_2{\cal J}){\cal D}_1\partial_{xx}\no\\
& & -6({\cal D}_1{\cal J}){\cal D}_2\partial_{xx}-6{\cal J}_x\partial_{xx}\no\\
& & + \biggl( 5T-2([{\cal D}_1,D_2]{\cal J})+\frac{8}{c}{\cal J}^2\biggr)
[D_1,D_2]\partial_x\no\\
& & -\biggl( 8({\cal D}_2{\cal J}_x)+\frac{16}{c}{\cal J}({\cal D}_2
{\cal J})+5({\cal D}_2T)\biggr){\cal D}_1\partial_x\no\\
& &+\biggl(-8({\cal D}_1{\cal J}_x)+\frac{16}{c}{\cal J}({\cal D}_1{\cal J})+5
(D_1T)\biggr){\cal D}_2\partial_x\no\\
& & +\biggl(\frac{3}{2}([D_1,D_2]T)-6{\cal J}_{xx}+u_3\biggr)\partial_x\no\\
& &-(3({\cal D}_2T_x)+3({\cal D}_2{\cal J}_{xx})+\Psi){\cal D}_1\\
& &+(3({\cal D}_1T_x)-3(D_1{\cal J}_{xx})-\bar{\Psi}){\cal D}_2\no\\
&+&\biggl(-2{\cal J}_{xxx}+([D_1,D_2]T_x+\frac{1}{2}u_{3x}+\frac{1}{2}
({\cal D}_1
\Psi)+\frac{1}{2}({\cal D}_2\Psi)-\frac{4}{c}([D_1,D_2]{\cal J}{\cal J}_x)
\biggr)\no
\ee
where $c$ is an arbitrary constant (central extension term) while
\be
\Psi &=&\frac{8}{c}\partial_x({\cal J}{\cal D}_2{\cal J})-\frac{72}{c}T
{\cal D}_2{\cal J}+\frac{36}{c}([{\cal D}_1,{\cal D}_2]{\cal J})
({\cal D}_2{\cal J})+\frac{8}{c}{\cal J}({\cal D}_2T)\no\\
& &-\frac{128}{c^2}{\cal J}^2({\cal D}_2{\cal J})+\frac{4}{c}{\cal J}_x
({\cal D}_2{\cal J})\no\\
\bar{\Psi}&=&-\frac{8}{c}\partial_x({\cal J}{\cal D}_1{\cal J})-\frac{72}{c}
T{\cal D}_1{\cal J}+\frac{36}{c}([{\cal D}_1,{\cal D}_2]{\cal J})
({\cal D}_1{\cal J})+\frac{8}{c}{\cal J}({\cal D}_1T)\no\\
& &-\frac{128}{c^2}{\cal J}^2({\cal D}_1{\cal J})-\frac{4}{c}{\cal J}_x
({\cal D}_1{\cal J})\no\\
u_3&=&\frac{56}{c}{\cal J}T-\frac{32}{c}{\cal J}([D_1,D_2]{\cal J})+\frac{128}
{c^2}{\cal J}^3+\frac{120}{c}({\cal D}_1{\cal J})({\cal D}_2{\cal J})\ .\no
\ee
Using this operator to the equation
$$\left(\begin{array}{c}{\cal
J}\\T\end{array}\right)_t=P_2\cdot\bigtriangledown
\int (T+\alpha{\cal J}^2)dx\ ,\eqno{(2.13)}$$
where $\alpha$ is an arbitrary constant, we obtain the $N=2$ supersymmetric
extension of the Boussinesq equation
\add
\renewcommand{\theequation}{\arabic{equation}.14}
\be
{\cal J}_t&=&2T_x+\alpha\biggl(\frac{c}{4}([D_1,D_2]{\cal J}_x)+4{\cal J}{\cal
J}
_x\biggr)\no\\
T_t&=&-2{\cal J}_{xxx}+([{\cal D}_1,{\cal D}_2]T_x)+\frac{80}{c}(D_1{\cal J}
D_2{\cal J})_x-\frac{16}{c}({\cal J}[D_1,D_2]{\cal J})_x\no\\
& &-\frac{16}{c}{\cal J}_x([D_1,D_2]{\cal J})+\frac{256}{c^2}{\cal J}^2{\cal
J}_x
+\biggl(\frac{40}{c}-2\alpha\biggr)({\cal D}_1{\cal J}{\cal D}_2T)\no\\
& & +\biggl(\frac{64}{c}+4\alpha\biggr){\cal J}_xT+\biggl(\frac{24}{c}+
2\alpha\biggr){\cal J}Tx+\biggl(\frac{40}{c}-2\alpha\biggr)({\cal D}_2{\cal J}
)(D_1T)\ .
\ee
Now we would like to check wheather the Hamiltonian operator $P_2$ produces,
similarly as in the classical case, the first Hamiltonian operator. Therefore
let us shift the $T\rightarrow T+\lambda$ in the $P_2$ operator, obtaining
\add
\renewcommand{\theequation}{\arabic{equation}.15}
\be
P_1=\left(\begin{array}{ccc}0&,&2\partial_x\\2\partial_x&,&
\left.\begin{array}{cc}5[D_1,D_2]\partial_x+\frac{1}{c}\{ 64{\cal J}_x+56{\cal
J}
\partial_x\\+72({\cal D}_2{\cal J}){\cal D}_1+72(D_1{\cal J})D_2\}\end{array}
\right.\end{array}\right)
\ee

It is easy to check that $P_1$ is an antisymmetric operator with the
vanishing Shouten bracket [24] and hence $P_1$ defines a proper Hamiltonian
operator.

In order to find the hamiltonian which produce (via $P_1$) the equation (2.14)
we show that such exist for $\alpha=-16/c$. Indeed let us first simplify the
equation 2.14 by shifting the $T$ superfunction to
$$T\rightarrow T+2[D_1,D_2]{\cal J}+16{\cal J}^2\eqno{(2.16)}$$
Then the super-Boussinesq equation for $\alpha=-16/c$ reduces to
$${\cal J}_t=2T_x\eqno{(2.17)}$$
$$T_t=-3([D_1,D_2]T_x)-\frac{72}{c}{\cal J}T_x+$$
$$+\frac{72}{c}[(D{\cal J})(D_2T)+(D_2{\cal J})(D_1{\cal J})]\eqno{(2.18)}$$
while the $P_1$ operator transform to the following form
$$P_1=\left(\begin{array}{ccc}0&,&2\partial_x\\
2\partial_x&,& \left.\begin{array}{ll}-3[D_1,D_2]\partial_x-\frac{72}{c}
{\cal J}\partial_x\\ +\frac{72}{c}[(D_1{\cal J}){\cal D}_2+({\cal D}_2
{\cal J}){\cal D}_1]\end{array}\right.\end{array}\right)\eqno{(2.19)}$$
Now it is not so diffucult to prove that the following quantity
$$H_1=\int\frac{1}{2}T^2dx$$
is the conserved quantity and produces  via (2.19) the equation (2.17-2.18).

Interestingly, for $\alpha=-\frac{16}{c}$ the supersymmetric equation (2.14)
is not reduced to the usual classical Boussinesq equation. This limit exists
only for $\alpha=-\frac{4}{c}$ [16]. Moreover our first hamiltonian is not
reduced in this limit  to the classical partner either. Thus our
bi-hamiltonian structure describes a pure supersymmetric effect.

Usually in the soliton's theory, the bihamiltonian structure is established
via the celebrated Adler-Konstant-Symes scheme [3,8]. In this construction we
have to know the Lax pair. In the next section we present several different
supersymmetric extensions of the Lax pair and we show that one of them
produces the supersymmetric extension of the Boussinesq equation and
coincides with the equation (2.14) only for $\alpha=-\frac{4}{c}$, e.g.
the one which contains the classical case.
\section{The supersymmetric Lax operator for the supersymmetric
Boussinesq equation.}

In order to find the supersymmetric extension of the Lax pair (1.3), first
let us consider the super-pseudo-differential elements of the supersymmetric
algebra ${\cal G}$
$${\cal G}\in q=\left\{\begin{array}{c}\sum\limits^\infty_{n=-\infty}
(b_n+f_n{\cal D}_1+ff_n{\cal D}_2+bb_nD_1D_2)\partial^n\end{array}\right\}
\ ,\eqno{(3.1)}$$
where $b_n,bb_n$ are the superbosons while $f_n,ff_n$ are the superfermions.
The action of the operator $\partial^{-1}$ is the formal integration defined
by the basic rule
$$\partial^{-1}b=b^{\partial^{-1}}-b_x\partial^{-2}+b_{xx}\partial^{-3}\ ,
\eqno{(3.2)}$$
$$b\partial^{-1}=\partial^{-1}b+\partial^{-2}b_x+\partial^{-3}b_{xx}\ ,
\eqno{(3.3)}$$
where $b$ is a some function.

Let us postulate the following most general form of the Lax operator to be
$$L=a\partial_{xxx}+\beta D_1D_2\partial_{xx}+Z1\partial_x+Z0\eqno{(3.4)}$$
where $a$ and $\beta$ are arbitrary constants and $Z1$ and $Z0$ are the
elements of the superalgebra ${\cal G}$. The elements $Z1$ and $Z0$ are
constructed
from all possible combinations of the elements ${\cal D}_1,{\cal D}_2,
\partial_x$ and two superboson fields $\cal J$ and $T$. There objects are
too complicated to be presented here. By using the symbolic manipulation
package REDUCE, we have verified and simplified this ansatz and finally
have found several different supersymmetric Boussinesq equations. The most
interesting is the one which is reduced to the equation (2.14). For this case
the Lax operator (3.4) take the following form
$$L={\cal D}_1[\partial_{xx}+\eta{\cal J}\partial_x+T]{\cal D}_2\eqno{(3.5)}$$
where $\eta$ is an arbitrary constant.

Substituting $L$ to the (1.4) we obtained the following equation
$${\cal J}_t=\frac{\gamma}{3\eta}[3\eta{\cal J}_x+\eta^2{\cal J}^2-6T]_x
\eqno{(3.6)}$$
\renewcommand{\theequation}{\arabic{equation}.7}
\be
T_t &=& \gamma\biggl[\frac{2}{3}\eta{\cal J}_{xxx}+\frac{2}{3}\eta^2{\cal J}
{\cal J}_{xx}+\frac{2}{3}\eta(T{\cal J}_x-{\cal J}T_x)\\
& & +\frac{2}{3}\eta^2(D_1{\cal J})(D_2{\cal J}_x)-T_{xx}+\frac{2}{3}
\eta[[(D_1T)(D_2{\cal J})+(D_2T)(D_1{\cal J})]\biggr]\ .\no
\ee
This equation coincides with the equation (2.14) only for $\alpha=-4/c,\
\gamma=3,\ \eta=24/c$. To see it let us shift the function $T$ in
(3.6-3.7) to
$$T\rightarrow -\frac{8}{c}T+\frac{12}{c}{\cal J}_xz+\frac{96}{c^2}{\cal J}^2
\ ,\eqno{(3.8)}$$
and next change
$${\cal D}_1\rightarrow{\cal D}_2\ \ ,\ \ \ {\cal D}_2\rightarrow{\cal D}_1
\ .\eqno{(3.9)}$$
As the result we obtained the following equation
$${\cal J}_t=2T_x\eqno{(3.10)}$$
$$T_t=-\frac{3}{2}{\cal J}_{xxx}+\frac{72}{c}((D_1{\cal J})(D_2{\cal J}))_x
+\frac{48}{c}{\cal J}_xT$$
$$+\frac{48}{c}[(D_1{\cal J})(D_2T)+(D_2{\cal J})(D_1T)]+\frac{576}{c^2}
{\cal J}^2{\cal J}_x\eqno{(3.11})$$
Now we make the similar transformation for the equation (2.14) e.q. shift $T$
to
$$T\rightarrow T+\frac{1}{2}([D_1,D_2]{\cal J})+\frac{4}{c}{\cal J}^2
\eqno{(3.12)}$$
and obtain that (2.14) is equivalent with (3.10-3.11).

Interestingly the transformation (3.9) give us also the new form of the
Lax operator
$$L={\cal D}_2\cdot[\partial_{xx}+\eta{\cal J}\cdot\partial_x+T]{\cal D}_1\ .
\eqno{(3.13)}$$
Thus the system described  by the equations (3.6-3.7) and by (3.10-3.11) for
$\alpha=-4/c$ is integrable and the conservation  laws are given by the
standard trace  form
$$I_n=tr L^{\frac{n}{3}}\eqno{(3.14)}$$
where $tr$ denotes the coefficient standing in the $\partial^{-1}{\cal D}_1D_2$
term.

For the completeness, let us present other Lax operator which produce also the
different supersymmetric extension of the Boussinesq equation. These Lax
pairs follows from the observation that the supersymmetric cubic roots of the
$L$ operator are not uniquelly defined\renewcommand{\thefootnote}{*}
\footnote{Notice that the same problem has appeared in the $N=2$ supersymmetric
KdV Lax pair where we have two different roots of the Lax pair [19]}.
Indeed one can quickly check that
$$(\varepsilon\partial_x+XD_1D_2)^3=a\partial_{xxx}+\beta D_1D_2\partial_{xx}
\eqno{(3.15)}$$
if
$$a=\epsilon^3\eqno{(3.16)}$$
$$\beta=X^3-3\epsilon X^2+3\epsilon X \eqno{(3.17)}$$
As we see from the last formulas, there are four different solutions of
$a,\beta,X,\epsilon$.

Namely, the first one characterize $a=0,\ \epsilon=0$ and therefore without
loosing any generality we can set $\beta=a=1$. This case has been considered
previously and produces the equations (3.6-3.7).

For the other cases we can set $\epsilon=1$ without loosing on the generality
of discussion as well. Then the equation (3.14) has three different solutions
$$X_1=\frac{1}{2}[i(\beta-1)^{\frac{1}{3}}\sqrt{3}-(\beta-1)^{\frac{1}{3}}+2]
\eqno{(3.18)}$$
$$X_2=\frac{1}{2}[i(\beta-1)^{\frac{1}{3}}\sqrt{3}+(\beta-1)^{\frac{1}{3}}-2]
\eqno{(3.19)}$$
$$X_3=(\beta-1)^{\frac{1}{3}}+1\eqno{(3.20)}$$
Let us denote by $L_i^{1/3}\ \ i=1,2,3$ the corresponding roots of the Lax
operator. Notice that two of the roots are complex, but it does not disturb
us in the construction of the Lax pair in the form (1.4) because we have to
construct $(L^{2/3})_+$ which could take real values. Interestingly $X^2_3$
and $X_1\cdot X_2$ are real numbers so in the general we should also consider
the
Lax pair in the form
$$L_{it}=\gamma[L_i,(L_i^{\frac{1}{3}}L_j^{\frac{1}{3}})_+]
\eqno{(3.21)}$$
However, we have known that $(L_i^{\frac{1}{3}}L_j^{\frac{1}{3}})_+=(L_j
^{\frac{1}{3}}L_i^{\frac{1}{3}})_+$ for all $i,j=1,2,3$. Therefore we have to
check the six different Lax pairs only. Using once more the symbolic
manipulation program REDUCE we found that for two Lax pairs only $(i=j=3\
{\rm and}\ i=1,j=2)$ the equation (3.21) produce the same real equation which
could be written down as
$${\cal J}_t=T_x \eqno{(3.22)}$$
$$T_t=a_0(a_1{\cal J}_{xx}+a_2{\cal J}^3+a_3(D_2J)(D_1{\cal J}))_x
\eqno{(3.23)}$$
$$+a_4[T{\cal J}_x+(D_2{\cal J})(D_1T)+(D_1{\cal J})(D_2T)]$$
where
$$
\begin{array}{lll}
a_0=3(\beta-1)^{\frac{1}{3}}\lambda &, & a_1=-3\beta^3+6\beta-3\ ,\\
a_2=\frac{2}{3}\eta^2 &, & a_3=6(1-\beta)\eta\ ,\\
a_4=2\lambda\eta &, & \gamma=\lambda^2\ ,\ \beta\neq 1\end{array}
\eqno{(3.24)}$$
and $\lambda$ is an arbitrary parameter.

The $L$ operator in that case takes the form
$$L=\partial_{xxx}+\beta D_1D_2\partial_{xx}+\eta {\cal D}_1\cdot{\cal J}\cdot
D_2+D_1\cdot TD_2
\eqno{(3.25)}$$
The case $\beta=1$ should be considered independently. Interestingly
for this case we have found the Lax pair in the form (3.13).
\vskip 2.0 cm

\section{Conclusion.}

As we see in this paper it is possible to construct the bihamiltonian structure
for the supersymmetric Boussinesq equation. However, our structure is not
reduced to the classical bihamiltonian structure and hence it describes a pure
supersymmetric effect.

For the superextension which contains the classical Boussinesq equation, we
have found the Lax operator. The knowledge of the Lax operator is an important
property which allows us to perform the investigation of the given system in
more detail on a deeper level. For example, applying the celebrated
Adler-Konstant-Symes scheme [3,8], it is possible, in principle, to find the
bihamiltonian system and hence the Poisson brackets for the given equation
using the Lax pair only. Applying this method to our supersymmetric Lax
operator, we can find the Poisson bracket but defined now on a larger
subspace than the one on which the Lax operator takes values. Therefore, we
expect that we have to apply the Dirac reduction technique in order to
find the Poisson bracket on the same subspace as the Lax operator. This problem
we postpone to a further communication.
\vskip 2.0 cm
{\bf Acknowledgements:} The author thanks the Research Institute for
Theoretical
Physics, University of Helsinki, for the hospitality.

The paper has been supported in parts by KBN grant 106/p3/g2/03.
\pagebreak

\end{document}